# ANALYSIS OF THE SECURITY OF BB84 BY MODEL CHECKING

Mohamed Elboukhari[1], Mostafa Azizi[2] and Abdelmalek Azizi[1,3]

[1]dept. Mathematics & Computer Science, FSO, University Mohamed I[st], Morocco
[2]dept. Applied Engineering, ESTO, University Mohamed I[st], Oujda, Morocco
elboukharimohamed@gmail.com , azizi.mos@gmail.com
[3]Academy Hassan II of Sciences & Technology, Rabat, Morocco
abdelmalekazizi@yahoo.fr

## ABSTRACT

*Quantum Cryptography or Quantum key distribution (QKD) is a technique that allows the secure distribution of a bit string, used as key in cryptographic protocols. When it was noted that quantum computers could break public key cryptosystems based on number theory extensive studies have been undertaken on QKD. Based on quantum mechanics, QKD offers unconditionally secure communication. Now, the progress of research in this field allows the anticipation of QKD to be available outside of laboratories within the next few years. Efforts are made to improve the performance and reliability of the implemented technologies. But several challenges remain despite this big progress. The task of how to test the apparatuses of QKD For example did not yet receive enough attention. These devises become complex and demand a big verification effort. In this paper we are interested in an approach based on the technique of probabilistic model checking for studying quantum information. Precisely, we use the PRISM tool to analyze the security of BB84 protocol and we are focused on the specific security property of eavesdropping detection. We show that this property is affected by the parameters of quantum channel and the power of eavesdropper.*

## 1. INTRODUCTION

The security has become a big task in wired and wireless networks. The characteristics of networks pose both challenges and opportunities in achieving security goals, such as confidentiality, authentication, integrity, availability, access control, and no repudiation. Cryptographic techniques are widely used for secure communications.

Classical cryptography is composed schematically by two systems: symmetric encryption and asymmetric encryption.

The cryptosystems of symmetric encryption use the same key for cipher and decipher messages. The key must be preserved secret by the parties of a communication. So in a network of people wanting to communicate in a confidential way with a cryptosystems of symmetric encryption, it is necessary that the keys are distinct. Precisely, it is necessary to create and distribute $n(n-1)/2$ keys which are distinct and secret. As we can remark, the cryptosystems of symmetric encryption suffer from the problem of creation and distribution the keys. This problem is mainly solved by the installation of the cryptosystems of asymmetric encryption [1].

A cryptosystem of asymmetric encryption operates by handling two keys: secret and public. Each participant diffuses a public key with his name. If one wishes to communicate with a participant, it is necessary to recover his public key and cipher with it the message, and send the ciphered message to this participant which is the only person who knows the secret key which makes possible to decipher the received messages. The secret key is of course related to the public key, in practice by a mathematical relation.

So, classical cryptography algorithms are based on mathematical functions. The robustness of a given cryptosystem is based essentially on the secrecy of its (private) key and the difficulty with which the inverse of its one-way function(s) can be calculated. Unfortunately, there is no mathematical proof that will establish whether it is not possible to find the inverse of a given one-way function. On the contrary, Quantum Cryptography (or Quantum Key Distribution QKD) is a method for sharing secret keys, whose security can be formally demonstrated. The use of Quantum Cryptography will enforce safety, dependability and security of information and communication technologies infrastructures and critical infrastructures.

QKD aims at exploiting the laws of quantum physics in order to carry out a cryptographic task. The idea of QKD did not attract much attention at first, research efforts have increased since the 1990s when it was proved that quantum computers could break the public-key cryptosystems commonly used in modern cryptography. Also a more interest has been generated after the first practical demonstration over 30 cm of free space employing polarisation coding [2]. Various theoretical and experimental studies have been undertaken, and prototype products are now commercially available.

The laws of quantum physics guarantee the security of quantum cryptography protocols. The BB84 protocol is the first quantum cryptography protocol, which was proposed by Bennett and Brassard in 1984 [2]. The security proof of this protocol against arbitrary eavesdropping strategies was first proved by Mayers [3], and a simple proof was later shown by Shor and Preskill [4].

The mathematical proof of security of quantum cryptography protocols is not enough to assure that the implementation of a system related to certain quantum protocol is secure. As in traditional cryptography, during the progress from an ideal protocol to an implementation, several flaws of security can appear. Thus, even extensive research has been initiated for sophisticated implementation of quantum cryptography in practical communication networks, these systems are difficult to design; it is very important to analyze and verify such systems with more details related to their practical implementation.

In our paper we present an analysis using PRISM [5]; a tool of the technique of probabilistic model checking. Our work is done in the same manner as [6] and [7], but our effort is concentrated on the property of detecting the eavesdropper. We also introduce new parameters of quantum channel's efficiency and the parameter of the eavesdropper's power. We show that these parameters affect the detecting of the eavesdropper.

The remainder of this article is organized as follows. The related works is introduced in section 2. In Section 3 we provide a detailed description of the BB84 protocol. In Section 4 we give a simple presentation of the technique of model checking and we show why this technique is desired to analyze protocols of quantum key distribution. In section 5 we present our analysis of BB84's security by introducing parameters of the channel and the eavesdropper in order to study the property of eavesdropping detection. Section 6 concludes our work by giving the main results.

## 2. RELATED WORKS

Analysis of quantum protocols is treated by certain authors. Especially the issue of analyzing quantum protocols by the technique of model checking is already introduced in the literature. More specially, using the approach of model checking for studying quantum cryptography protocols has been also evoked.

In the article [8], the authors Rajagopal Nagarajan and Simon Gay propose to analyze quantum protocols by the techniques of formal verification which was applied and developed in classical computing for the analysis of communicating concurrent systems. In such techniques, the first step in formal verification is to define a model of the system to be analysed, in a well-founded mathematical notation. Next, based on the same underlying theory, an automated analysis tool is used to reason about the system.

The authors Rajagopal Nagarajan, Simon Gay and Nikolaos Papanikolaou In their article [9] describe fundamental and general techniques for formal verification of quantum protocols. Because current analyses of quantum protocols use a traditional mathematical approach and require considerable understanding of the underlying physics, the authors argue that automated verification techniques provide an elegant alternative. They show the feasibility of these techniques through the use of PRISM, a probabilistic model-checking tool. In their articles [10]-[11], they establish model-checking techniques for the automated analysis of quantum information protocols. Precisely they have described QMC, a model-checking tool for quantum protocols. As opposed to simulation systems, QMC is the first dedicated verification tool for quantum protocols. QMC enables the modeling and verification of properties of quantum protocols expressible in the quantum formalism.

Directed related to our work, in the article [12] the authors Rajagopal Nagarajan, Nikolaos Papanikolaou, Garry Bowen and Simon Gay introduce the use of computer–aided verification as a practical means for analysing the QKD protocol BB84. based on probabilistic model–checking approach, they have used the PRISM model–checker to show that, the equivocation of the eavesdropper with respect to the channel decreases exponentially as the number of qubits transmitted in BB84 is increased. They have also shown that the probability of detecting the presence of an eavesdropper increases exponentially when the number of qubits.

In the article [7] the authors Mohamed Elboukhari, Mostafa Azizi, and Abdelmalek Azizi present a methodology based on model checking for analyzing quantum information systems. Particularly they are interested in the QKD protocol B92. Using the PRISM tool as a probabilistic model checker, they demonstrate that the protocol B92 fulfilled specific security properties. The authors in the article [13] use the same technique to analyze certain security's properties of B92 protocol; they are interested in the specific security property of eavesdropping detection. They have shown that this property is affected by the power of eavesdropper and the parameters of quantum channel.

Others works related to the analysis of quantum protocols can be found in [6]-[14]-[15]-[16].

## 3. BB84 PROTOCOL

Quantum cryptography is only used to produce and distribute a key $K = \{0,1\}^N$, not to transmit any message data. This key can then be used with any chosen encryption algorithm to encrypt (and decrypt) a message, which can then be transmitted over a standard communication channel (classical channel).

The security of quantum cryptography relies on the foundations of quantum mechanics, in contrast to traditional public key cryptography which relies on the computational difficulty of certain mathematical functions. Also traditional public key cryptography cannot provide any indication of eavesdropping or guarantee of key security. Quantum key distribution has an important and unique properly; it is the ability of the two communicating users (traditionally referred to as Alice and Bob) to detect the presence of any third party (referred to as Eve) trying to gain knowledge of the key. A third party trying to eavesdrop on the key must in some way measure it, thus introducing detectable anomalies. By using quantum superpositions or quantum entanglement and transmitting information in quantum states over a quantum channel (such as an optical fiber or free air), a communication system can be implemented which detects eavesdropping.

BB84 was the first studied and practical implemented QKD physical layer protocol. It was elaborated by Charles Bennet and Gilles Brassard in 1984 in their article [2]. It is surely the most famous and most realized quantum cryptography protocol. This scheme uses the transmission of single polarized photons (as the quantum states). The polarizations of the photons are four, and are grouped together in two different non orthogonal basis.

Generally the two non orthogonal basis are:
-base $\oplus$ of the horizontal (0°) and vertical polarization (+90°), and we represent the base states with the intuitive notation: $|0\rangle$ and $|1\rangle$. We have $\oplus = \{|0\rangle, |1\rangle\}$.

-base $\otimes$ of the diagonal polarizations (+45°) and (+135°). The two different base states are $|+\rangle$ and $|-\rangle$ with $|+\rangle = \frac{1}{\sqrt{2}}(|0\rangle + |1\rangle)$ and $|-\rangle = \frac{1}{\sqrt{2}}(|0\rangle - |1\rangle)$. We have $\otimes = \{|+\rangle, |-\rangle\}$.

In this protocol, the association between the information bit (taken from a random number generator) and the basis are described in Table 1.

Table 1. Coding scheme for the BB84 protocol.

| Bit | $\oplus$ | $\otimes$ |
|---|---|---|
| 0 | $|0\rangle = a_{00}$ | $|+\rangle = a_{10}$ |
| 1 | $|1\rangle = a_{01}$ | $|-\rangle = a_{11}$ |

The BB84 can be described as follows [17]:
1) Quantum Transmissions (First Phase)
   a) Alice chooses a random string of bits $d \in \{0,1\}^n$, and a random string of bases $b \in \{\oplus, \otimes\}^n$, where $n > N$.
   b) Alice prepares a photon in quantum state $a_{ij}$ for each bit $d_i$ in $d$ and $b_i$ in $b$ as in Table 1, and sends it to Bob over the quantum channel.
   c) With respect to either $\oplus$ or $\otimes$, chosen at random, Bob measures each $a_{ij}$ received. Bob's measurements produce a string $d' \in \{0,1\}^n$, while his choices of bases form $b' \in \{0,1\}^n$.
2) Public Discussion (Second Phase)
   a) For each bit $d_i$ in $d$
      i) Alice over the classical channel sends the value of $b_i$ to Bob.
      ii) Bob responds to Alice by stating whether he used the same basis for the measurement. Both $d_i$ and $d'_i$ are discarded if $b_i \neq b'_i$.
   b) Alice chooses a random subset of the remaining bits in $d$ and discloses their values to Bob over the classical channel (over internet for example). If the result of Bob's measurements for any of these bits does not match the values disclosed, eavesdropping is detected and communication is aborted.
   c) the string of bits remaining in $d$ once the bits disclosed in step *2b)* are removed is the common secret key, $K = \{0,1\}^N$.

To understand BB84 protocol it very important to describe how we measure a qubit in the field of quantum physics; if we have a qubit as $|qubit\rangle = e|c\rangle + f|g\rangle$ so the measure of this state in the basis $\{|c\rangle, |g\rangle\}$ produces the state $|c\rangle$ with the probability of $|e|^2$ and the state of $|g\rangle$ with the probability of $|f|^2$ and of course $|e|^2 + |f|^2 = 1$ ($|e|^2$ is the absolute square of the amplitude of $e$). So, measuring with the incorrect basis yields a random result, as predicted by quantum theory. Thus, if Bob chooses the $\otimes$ basis to measure a photon in state $|1\rangle$, the classical outcome will be either 0 or 1 with equal probability because $|1\rangle = \frac{1}{\sqrt{2}}(|+\rangle - |-\rangle)$; if the $\oplus$ basis was chosen instead, the classical outcome would be 1 with certainty because $|1\rangle = 1|1\rangle + 0|0\rangle$.

To detect Eve, Alice and Bob perform a test for eavesdropping in step *2b)* of the protocol. The idea is that, wherever Alice and Bob's bases are identical (i.e. $b_i = b_i^{'}$), the corresponding bits should match (i.e. $d_i = d_i^{'}$). If not, an external disturbance is produced or there is noise in the quantum channel, we suppose all that is caused by Eve. In our article we are interested in analyzing this important property assured by quantum mechanics: the enemy's presence is always made manifest to the legitimate users.

## 4. WHY MODEL CHECKING FOR QUANTUM KEY DISTRIBUTION?

In software and hardware design of complex systems, more time and effort are spent on verification than on construction. Techniques are sought to reduce and ease the verification efforts while increasing their coverage. In this context, formal verification is the act of proving or disproving the correctness of intended algorithms underlying a system with respect to a certain formal specification or property, using formal methods of mathematics. Model checking is an approach of formal verification. Model checking is a verification technique that explores all possible system states in a brute-force manner. In the field of logic in computer science, model checking refers to the following problem: Given a model of a system, test automatically whether this model meets a given specification. Using a specialized software tool (called a model–checker), a system implementor can mechanically prove that the system satisfies a certain set of requirements.

In the literature there are several Proofs of unconditional security of the BB84 protocol [3]-[4], but as Gottesman and Lo [18] point out that "the proof of security of QKD is a fine theoretical result, but it does not mean that a real QKD system would be secure". Thus more flexible approach to analyzing the security of quantum cryptographic protocols is clearly desirable. A real component of a system may be quantum, but others could still be classical. So, manufacturers of commercial quantum cryptographic systems [19], for instance, require efficient and rigorous methods for design and testing.

In our article we propose to analyze the security of BB84 protocol by model checking. Generally to realize this, we build an abstract model, noted *M* and we express it in a description language. Also, we describe the desired behavior of the system in a set of temporal formulae $p_i$. The model and the formulae are the input of the model–checker.

For systems which have a probabilistic behavior, a variation of this technique is used; a probabilistic model–checker, such as PRISM [20]. System properties for PRISM models are written in Probabilistic Computation Tree Logic (PCTL). PRISM models are represented by probabilistic transition systems.

In PRISM we verify if the model $M$ satisfy the property defined by $p_i$ (i.e. whether for each $M \vDash p_i$ property $p_i$), and with PRISM we compute the probability:

$$P_r\{M \vDash p_i\} \quad (1)$$

Also, we can parameterize the model $M$ by writing $M = M(x_1, x_2, x_3, \ldots, x_n)$ and the probability (1) can be calculated for different value of $x_i$, this enables us to have a meaningful plot of the variation of (1).

A model in PRISM is formed by components called modules. Each module has a sequence of actions to be achieved and its own local variables. The actions take the following form:

$$[action] \to a_1 : (\text{var}_1 = value_1) + a_2 : (\text{var}_2 = value_2) + \ldots + a_n : (\text{var}_n = value_n) \quad (2)$$

In this equation, the variable $\text{var}_i$ is assigned by $value_i$ with probability $a_i$ ($\sum_{i=1}^{i=n} a_i = 1$). In case where $n = 1$ we have the notation: $a_1 : (\text{var}_1 = value_1) = (\text{var}_1 = value_1)$ with $a_1 = 1$. The model checker PRISM permits us to specify arbitrarily probabilities for actions, for example in case $n = 2$ we can model a tendency in BB84 protocol of Alice in the choice of the quantum states by a module containing the action:

$$[EtatOfAlice] true \to 0.8 : (EtatAlice = |1\rangle) + 0.2 : (EtatAlice = |0\rangle); \quad (3)$$

Here, Alice is biased towards choosing the state $|1\rangle$ to encode the data 1 as in Table 1.

## 5. ANALYSIS OF BB84 PROTOCOL USING THE MODEL CHECKER PRISM

### 5.1 Description of the Model BB84 in PRISM Tool

PRISM is a probabilistic model checking tool being developed at the University of Birmingham. Conventional model checkers input a description of a model, represented as a state transition system, and a specification, typically a formula in some temporal logic, and return "yes" or "no", indicating whether or not the model satisfies the specification. In the case of probabilistic model checking, the models are probabilistic, in the sense that they encode the probability of making a transition between states instead of simply the existence of such a transition, and analysis normally entails calculation of the actual likelihoods through appropriate numerical or analytical methods.

We have elaborated a model of BB84 in PRISM noted $M_{BB84}$. It is done within a file containing modules that represent the components of the system. In our model of BB84, there is a module corresponding to each party involved in the protocol (Alice, Bob and Eve), plus a module representing the quantum channel.

As mentioned before, we are interested to studying by PRISM the specific security property that BB84 protocol must offer: an enemy who tries to eavesdrop must be detected. So, as mentioned in [2], if Alice and Bob know that Eve is trying to eavesdrop, they can be in agreement to use the technique of purification and/or temporarily to stop the key establishment process.

By using our model of BB84, we can calculate the probability:

$$P_r\{M_{BB84} \vDash p_{det}\} \quad (4)$$

Where $p_{det}$ represents a formula PCTL, its Boolean value is *TRUE* if the enemy is detected. We can vary $n$, the number of photons transmitted involved in the communication between Alice and Bob, and so in our PRISM model this probability is a function of $n$. Let us write the probability of detecting the enemy like:

$$P_{det}(n) = P_r\{M_{BB84} \vDash p_{det}\} \quad (5)$$

PRISM calculates exactly the probability of detecting an enemy, Eve. But we must give the definition of $p_{det}$. For that we must state the random event $\omega$ occurs when an enemy is detected, this will enable us to write $P_{det}(n)$ like a classical probability $P_r(\omega)$.

## 5.2. The Expression of $p_{det}$

In our model $M_{BB84}$, Eve performs the attack "main in the middle"; Eve receives each photon on the quantum channel, measures it with her basis denoted $b_i^"$, obtaining a bit value $d_i^"$, and then transmits to Bob a new photon, representing $d_i^"$ in the $b_i^"$ basis. Eve has to make a random choice of basis, denoted $b_i^"$, which may or may not match Alice's original choice, $b_i$. In the case $b_i^" = b_i$, Eve is sure that she measured the i-th photon correctly; otherwise, quantum theory predicts that her measurement result will be correct with the probability of $1/2$.

For the detection of Eve, Bob must choose the correct basis (as Alice) for his measurement. If Bob obtains an incorrect bit value despite having used the correct basis, this is because an enemy has caused a disturbance. Here we assume that the quantum channel is noisy and for the need of security we suppose all noise is due to Eve. So, Eve is detected as soon as the following event has occurs:

$$\omega = (b_i = b_i^') \wedge (d_i \neq d_i^') \text{ for some } i \leq n \quad (6)$$

Therefore, we can give the expression of $P_{det}(n)$ as a classical probability:

$$P_{det}(n) = P_r\{(b_i = b_i^') \wedge (d_i \neq d_i^') \text{ for some } i \leq n\} \quad (7)$$

The PCTL formula $p_{det}$ corresponding is:

$$p_{det} = \{TRUE \cup (b_i = b_i^') \wedge (d_i \neq d_i^')\} \quad (8)$$

## 5.3. Influence of Quantum Channel's Efficiency on the Eve's Detection

The quantum channel in our model BB84 is written in a module called *Quantum Channel* which can be in reality optical fiber or free air. We propose to simulate the influence of the channel's efficiency on the detection of Eve. For that we have elaborated three curves; curve where the channel is perfect and where it produces some noise and a lot of noise. We expect that where the channel becomes noisy, the probability of detecting Eve increases.

In the perfect quantum channel, there is no noise, we model this in the module *Quanum Channel* by the line:

$$[aliceput](ch\_state = 0) \rightarrow (ch\_state' = 1) \& (ch\_bas' = al\_bas) \& (ch\_bit' = al\_bit); \quad (9)$$

We use $ch\_state$, $ch\_bas$, $ch\_bit$, $al\_bas$, $al\_bit$ for respectively state, base and bit of the channel and base and bit of Alice. This line shows that the information sent by Alice (base and bit) remain unchanged before it received by Eve.

For $1 \leq n \leq 50$, PRISM calculates $P_{det}(n)$, this produces the curve of $P_{det}$ (noted $P_{det}^{ch(0)}$) as in Fig. 1.

To elaborate a curve of $P_{det}$ (noted $P_{det}^{ch(1)}$) where there is a bit noise in the channel; we change the line (9) by this:

$$[aliceput](ch\_state = 0) \rightarrow 0.7 : (ch\_state' = 1) \& (ch\_bas' = al\_bas) \& (ch\_bit' = al\_bit)$$
$$+ 0.1 : (ch\_state' = 1) \& (ch\_bas' = 1 - al\_bas) \& (ch\_bit' = al\_bit)$$
$$+ 0.1 : (ch\_state' = 1) \& (ch\_bas' = al\_bas) \& (ch\_bit' = 1 - al\_bit)$$
$$+ 0.1 : (ch\_state' = 1) \& (ch\_bas' = 1 - al\_bas) \& (ch\_bit' = 1 - al\_bit); \quad (10)$$

As we remark, the information of Alice has been changed in a little way. If we modify the line (9) by the following:

$$[aliceput](ch\_state = 0) \rightarrow 0.4 : (ch\_state' = 1) \& (ch\_bas' = al\_bas) \& (ch\_bit' = al\_bit)$$
$$+ 0.2 : (ch\_state' = 1) \& (ch\_bas' = 1 - al\_bas) \& (ch\_bit' = al\_bit)$$
$$+ 0.2 : (ch\_state' = 1) \& (ch\_bas' = al\_bas) \& (ch\_bit' = 1 - al\_bit)$$
$$+ 0.2 : (ch\_state' = 1) \& (ch\_bas' = 1 - al\_bas) \& (ch\_bit' = 1 - al\_bit); \quad (11)$$

We simulate a channel very noisy. This enables us to perform a curve of $P_{det}$ (noted $P_{det}^{ch(2)}$). The curves $P_{det}^{ch(1)}$ and $P_{det}^{ch(2)}$ are shown in Fig. 1.

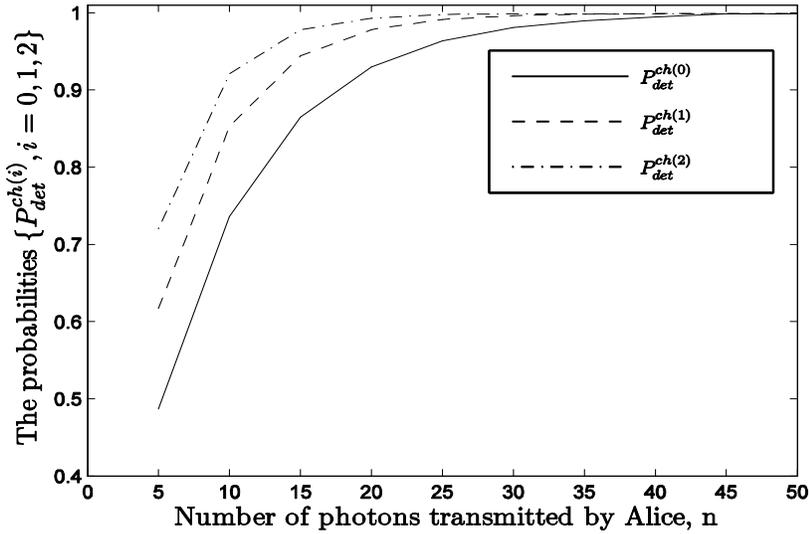

Figure 1. The probabilities $\{P_{det}^{ch(i)}(n), i = 0, 1, 2\}$ to detect Eve where the number of photons transmitted by Alice is between 5 and 50.

From the curves, we note that if we increase the number of photons emitted by Alice, the probability of detection of Eve increases and tends towards 1 and we have $\lim_{n \to \infty} P_{det}^{ch(i)} = 1$ for $0 \leq i \leq 2$. So, the property of detecting Eve in the BB84 protocol is clearly checked. Also, as the channel becomes noisy the probability of Eve's detection becomes higher as expected and we have the following inequality for $5 \leq n \leq 50$:

$$P_{det}^{ch(0)}(n) \leq P_{det}^{ch(1)}(n) \leq P_{det}^{ch(2)}(n) \quad (12)$$

## 5.4. Influence of Eve's Power of on its Detection

As in paragraph 5.3) we want to simulate the power of Eve and we expect that if the power is lower, the detection of Eve is lower too. In this paragraph, we suppose the quantum channel is perfect.

Firstly, the curve $P_{det}$ (noted also $P_{det}^{Eve(2)}$) represents the function $n \to P_{det}(n)$ where Eve is powerful; Eve performs the attack "man in the middle" for all photons emitted by Alice to Bob. This appears in the module *Quantum Channel* as:

$$[eveput](ch\_state = 3) \to (ch\_state' = 4) \& (ch\_bas' = eve\_bas) \& (ch\_bit' = eve\_bit); \quad (13)$$

Here, $eve\_bas$ and $eve\_bit$ refer to base and bit of Eve. If we write instead the line:

$$[eveput](ch\_state = 3) \to 0.2 : (ch\_state' = 4) \& (ch\_bas' = eve\_bas) \& (ch\_bit' = eve\_bit)$$
$$+ 0.8 : (ch\_state' = 4) \& (ch\_bas' = al\_bas) \& (ch\_bit' = al\_bit); \quad (14)$$

We model a weak attack of Eve because for several photons Eve doesn't intercept them. By varying $n$ in the interval [5, 70] we realize the curve of $P_{det}$; we note it by $P_{det}^{Eve(0)}$.

When Eve intercepts a lot of photons, we simulate a medium attack of Eve; this is done by changing the line (13) by the following:

$$[eveput](ch\_state = 3) \to 0.5 : (ch\_state' = 4) \& (ch\_bas' = eve\_bas) \& (ch\_bit' = eve\_bit)$$
$$+ 0.5 : (ch\_state' = 4) \& (ch\_bas' = al\_bas) \& (ch\_bit' = al\_bit); \quad (15)$$

In this situation PRISM provides a curve of $P_{det}$ noted $P_{det}^{Eve(1)}$. The curves $P_{det}^{Eve(i)}$, for $i = 0,1,2$ are illustrated in Fig. 2.

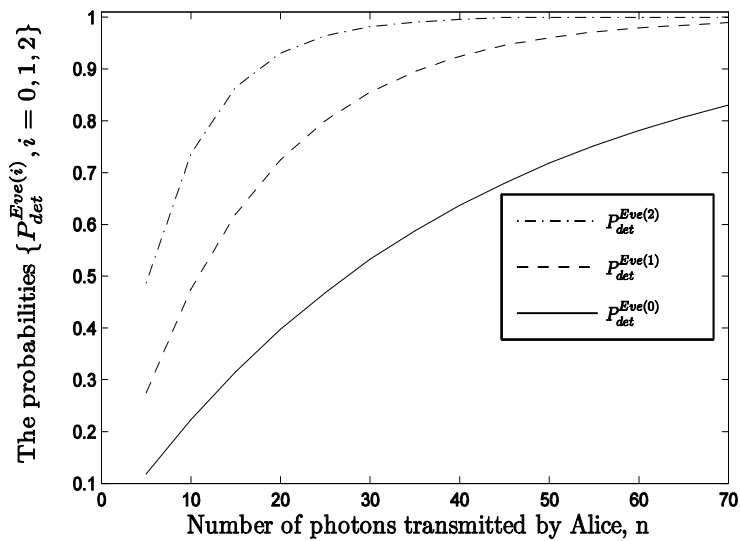

Figure 2. The probabilities $\{P_{det}^{Eve(i)}(n), i = 0, 1, 2\}$ to detect Eve where the number of photons transmitted by Alice is between 5 and 70.

In this figure we remark also that if we increase $n$, the number of photons transmitted by Alice, the probability of detection increase too and we have $\lim_{n \to +\infty} P_{\det}^{Eve(i)}(n) = 1$, $i = 0, 1, 2$ and thus the property of detection in BB84 protocol is verified.

More interesting, if the power of Eve become lower the probability of her detection becomes smaller, this clearly showed by the inequality for $5 \leq n \leq 70$ :

$$P_{\det}^{Eve(o)}(n) \leq P_{\det}^{Eve(1)}(n) \leq P_{\det}^{Eve(2)}(n) \tag{16}$$

## 6. CONCLUSION

Conventional cryptography such as symmetric and asymmetric cryptography, often involve the use of cryptographic keys. But all cryptographic techniques will be ineffective if the key distribution mechanism is weak. The security of these mechanisms of key distribution mechanism is based on computational complexity and the extraordinary time needed to break the code. QKD is attracting much attention as a solution of the problem of key distribution; QKD offers unconditionally secure communication based on quantum mechanics. And QKD could well be the first application of quantum mechanics at the single quanta level. Actually, many experiments have demonstrated that keys can be exchanged over distances of a few tens of kilometers at rates at least of the order of a thousand bits per second and there is no doubt that the technology can be mastered and will find commercial applications.

So, the QKD cryptosystems are very promising and the technology is improving more and more to fulfill requirements. Now, there is a big need of testing and analysis of such systems due to their complexity.

In this context, we have adopted the technique of model checking to analyze the security of the BB84 protocol. We have focused on studying the property of detecting the eavesdropper. By using the model checker PRISM we have the following results:

- To increase the probability of the detection of eavesdropper, it is necessary to increase the number of photons transmitted,

- If the quantum channel is noisy than the probability of detecting the eavesdropper increases too,

- If the power of Eve becomes stronger, the probability of her detection is higher.

So, the automatic model checker PRISM enables us to analyze BB84 protocol and this approach is adaptable to other protocol of quantum cryptography. Also it can be used to analyze heterogonous cryptographic systems containing classical and quantum components.

**Authors:**

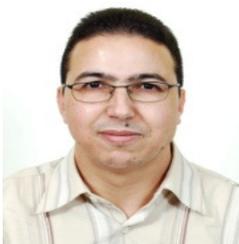

**Mohamed elboukhari** received the DESA (diploma of high study) degree in numerical analysis, computer science and treatment of signal in 2005 from the University of Science, Oujda, Morocco. He is currently a PhD student in the University of Oujda in the field of computer science. His research interests include cryptography, quantum cryptography and wireless network security.

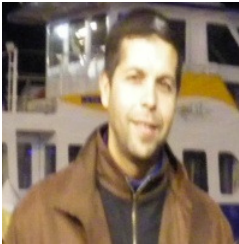

**Mostafa azizi** received the diploma of engineer in automatic and computer industry in 1993 from school Mohammadia of engineers, Rabat, Morocco. He received the Ph. D in computer science in 2001 from the university Montreal, Canada. He is currently professor at university of Mohamed first, Oujda, Morocco. His main interests include aspect of real time, embedded system, security and communication and management of the computer systems in relation with process industry.

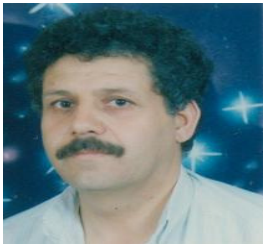

**Abdelmalek azizi** received the Ph. D in theory of numbers in 1993 from university Laval, Canada. He is professor at department of mathematics in university Mohamed first, Oujda, Morocco. He is interested in history of mathematics in Morocco and in the application of the theory of number in cryptography.